\documentclass[aps,preprint]{revtex4}

\usepackage{graphicx}

\newcommand{\be}{\begin{equation}}
\newcommand{\ee}{\end{equation}}

\begin{document}

\title{\bf LOCV calculations for polarized liquid $^3\mathrm{He}$ with the
spin-dependent correlation}
\author{G.H. Bordbar\footnote{Corresponding author}
\footnote{E-Mail: bordbar@physics.susc.ac.ir} and M.J. Karimi }
 \affiliation{
Department of Physics, Shiraz University, Shiraz 71454, Iran}

\begin{abstract}
We have used the lowest order constrained variational (LOCV)
method to calculate some ground state properties of polarized
liquid $^{3}He$ at zero temperature with the spin-dependent
correlation function employing the Lennard-Jones and Aziz pair
potentials. We have seen that the total energy of polarized liquid
$^{3}He$ increases by increasing polarization. For all
polarizations, it is shown that the total energy in the
spin-dependent case is lower than the spin-independent case. We
have seen that the difference between the energies of
spin-dependent and spin-independent cases decreases by increasing
polarization. We have shown that the main contribution of the
potential energy comes from the spin-triplet state.
\end{abstract}
 \maketitle

\section{Introduction}\label{S:intro}
Helium has two stable isotopes: that of mass 4 is readily
available as helium gas or liquid from the atmosphere or gas
wells, while that of mass 3 is extremely rare in nature and only
became available commercially in the 1950s \cite{Dob}. Liquid
$^{3}He$ is particularly suited to study correlation among the
strongly interacting many-body fermionic systems. Several
approaches have been used for investigating the properties of
normal liquid $^{3}He$. These are mainly based on the STLS scheme
\cite{Nd}, mott localization \cite{Gl}, spin fluctuation theory
\cite{Ms}, Green's function Monte Carlo (GFMC) \cite{GFMCa},
FN-DMC, DMC, VMC and EMC simulations \cite{DVEF}, CBPT formalism
\cite{FPS}, nonperturbative renormalization group equation
\cite{KW}, nonlocal density functional formalism \cite{PT} ,
correlated basis functions (CBF) \cite{CBFa} and Fermi
hyper-netted chain (FHNC) \cite{FHNCa}. The spin polarized liquid
$^{3}He$ as an interesting many-body system has been investigated
using different approaches such as  FHNC \cite{FHNCb}, GFMC
\cite{GFMCb}, CBF \cite{CBFb} and transport theory \cite{Trans}.

In recent years, we have studied both normal and polarized liquid
$^{3}He$ at zero and finite temperature
\cite{Gha1,Gha2,Gha3,Gha4}. In these calculations, the lowest
order constrained variational (LOCV) method based on the cluster
expansion of the energy functional has been used. This method is
fully self-consistent, since it does not introduce any free
parameter to the calculations. We have also used the LOCV method
in many-body calculations of dense matter \cite{Ghb}. Recently, we
have used this method to calculate some properties of the
polarized neutron matter and the polarized symmetrical and
asymmetrical nuclear matters \cite{Ghc}. In these works, a
comparison of our results and those of other many-body techniques
indicates that the LOCV method is a powerful microscopic technique
to calculate the properties of the polarized matter.

In this work, we use the LOCV method to compute the ground state
energy of the polarized liquid $^{3}He$ at zero temperature by
employing the spin-dependent correlation function with the
Lennard-Jones \cite{LJ} and Aziz \cite{Aziz1,Aziz2} pair
potentials.

\section{Lowest Order Constrained Variational Method\label{Met}}
We consider a system of $N$ interacting $^{3}He$ atoms with
$N^{+}$ spin up and  $N^{-}$ spin down atoms. The total number
density ($\rho$) and spin asymmetry parameter ($\xi$) are defined
as
\begin{eqnarray}
\rho&=&\rho^{+}+\rho^{-},\nonumber\\
\xi&=&\frac{N^{+}-N^{-}}{N}.
\end{eqnarray}
$\xi$ shows the spin ordering of matter which can have a value in
the range of $\xi=0.0$ (unpolarized matter) to $\xi=1.0$ (fully
polarized matter). For this system, we consider the energy per
particle up to the two-body term in the cluster expansion,
\begin{equation}
E=E_{1}+E_{2},
\end{equation}
where
\begin{eqnarray}
E_{1}&=&\frac{3}{10}\frac{\hbar^{2}}{m}(3\pi^{2}\rho)^{\frac{2}{3}}[(1+\xi)^{\frac{5}{3}}
+(1-\xi)^{\frac{5}{3}}],\nonumber\\
E_{2}&=&\frac{1}{2N} \sum_{i,j}\langle ij\mid w(12) \mid
 ij-ji\rangle.
\end{eqnarray}
 In the above equation, $w(12)$ is the effective pair potential,
\begin{equation}\label{Eff-pot}
 w(12)=-\frac{\hbar^{2}}{2m}[F(12),[\nabla_{12}^{2},F(12)]]+F(12)V(12)F(12),
\end{equation}
 where $F(12)$ is the two-body correlation
 operator and $V(12)$ is the pair potential between the helium atoms.
 In our calculations, we
use the Lennard-Jones \cite{LJ} and Aziz \cite{Aziz1,Aziz2} pair
potentials.
The Lennard-Jones pair potential is as follows \cite{LJ},
\begin{equation}
V(r) = 4\epsilon\left[\left(\frac{\sigma}{r}\right)^{12} -
\left(\frac{\sigma}{r}\right)^{6} \right],
\end{equation}
where
\begin{equation}
\epsilon = 10.22 K, \hskip 2truecm \sigma = 2.556 A\cdot
\end{equation}
The Aziz pair potential has the following form \cite{Aziz1,Aziz2},
\begin{equation}
V(r) = \epsilon\left\{ Ae^{-\alpha r/r_m} - \left[
C_6\left(\frac{r_m}{r} \right)^6 + C_8\left(\frac{r_m}{r}
\right)^8 + C_{10}\left(\frac{r_m}{r} \right)^{10}\right] F(r)
\right\},
\end{equation}
where
\begin{eqnarray}
F(r) & = & \left\{
\begin{tabular}{lll}
$e^{-(\frac{Dr_m}{r} - 1)^2}$
&;&$\frac{r}{r_m} \leq D$\\
1 &;&$\frac{r}{r_m} > D$,
\end{tabular}
\right.
\end{eqnarray}
and
\begin{eqnarray}
\begin{tabular}{ccc}
$\frac{\epsilon}{k_B} = 10.8 K$,& &$A = 0.5448504 \times 10^6$,\\
$\alpha = 13.353384$,& &$r_m = 2.9673 A$,\\
$C_6 = 1.37732412$,& &$C_8 = 0.4253785$,\\
$C_{10} = 0.178100$,& &$D = 1.241314 \cdot$
\end{tabular}
\end{eqnarray}

 Now, we consider a spin-dependent correlation function as follows
\begin{equation}\label{Corrf}
 F(12)=f_{0}(r_{12})P_{0}+f_{1}(r_{12})P_{1},
\end{equation}
where
\begin{eqnarray}
P_{0}&=&\frac{1}{4}(1-\sigma_{1}.\sigma_{2}),\nonumber\\
P_{1}&=&\frac{1}{4}(3+\sigma_{1}.\sigma_{2}).
\end{eqnarray}
 $f_{0}$ and $f_{1}$  indicate the spin-singlet and spin-triplet two-body correlation
functions, respectively. With the above two-body correlation
function, we have derived the following relation for the effective
pair potential,
\begin{equation}\label{W_{r12}}
w_{s}(r)=\frac{\hbar^{2}}{m}({f_{s}{'}}(r))^{2}+f_{s}^{2}(r)V(r),
\end{equation}
and then the two-body energy $E_{2}$ is found by
\begin{equation}\label{E_{2}}
 E_{2}=2\pi\rho\sum_{s=0,1}\int_{0}^{\infty}
drr^{2}w_{s}(r)a_{s}.
\end{equation}
In Eq. (\ref{E_{2}})
\begin{eqnarray}
a_{0}&=&\frac{1}{4}(1-\xi^{2})[1+l(k_{F^{+}}r)l(k_{F^{-}}r)],\nonumber\\
a_{1}&=&\frac{1}{4}[(1+\xi)^{2}(1-l^{2}(k_{F^{+}}r))
+(1-\xi)^{2}(1-l^{2}(k_{F^{-}}r))\nonumber\\
&&+(1-\xi^{2})(1-l(k_{F^{+}}r)l(k_{F^{-}}r))].
\end{eqnarray}
$k_{F^{\pm}}=(6\pi^{2}\rho^{\pm})^{\frac{1}{3}}$
 is the Fermi momentum and $l(x)$ is given by
\begin{equation}
l(x)=\frac{3}{x^{3}}[\sin(x)-x\cos(x)].
\end{equation}

Now, we minimize the two-body energy Eq. (\ref{E_{2}}) with
respect to the variations in the two-body correlation function
subject to the normalization constraint \cite{Clark},
\begin{equation}
\frac{1}{N}\sum_{i,j}\langle ij\vert h^{2}(12)-F^{2}(12)\vert
ij-ji\rangle=1.
\end{equation}
The normalization constraint is conveniently re-written in the
integral form as
\begin{equation}\label{Cons}
 4\pi\rho \sum_{s=0,1}\int_{0}^{\infty}
drr^{2}[h^{2}(r)-f_{s}^{2}(r)]a_{s}=1,
\end{equation}
 where the Pauli function $h(r)$ is
\begin{equation}
 h(r)=\{1-\frac{1}{4}[(1+\xi)^{2}l^{2}(k_{F^{+}}r)
+(1-\xi)^{2}l^{2}(k_{F^{-}}r)]\}^{-\frac{1}{2}}.
\end{equation}
The minimization of the two body energy $E_{2}$  gives the
following Euler-Lagrange differential equation for the two-body
correlation function $f_{s}(r)$,
\begin{equation}\label{Dif}
f_{s}^{''}(r)+(\frac{2}{r}+\frac{a_{s}^{'}}{a_{s}})f_{s}^{'}(r)-\frac{m}
{\hbar^{2}}\left(V(r)-2\lambda\right)=0.
\end{equation}
The Lagrange multiplier $\lambda$ imposes by normalization
constraint. For $s=0$ and $s=1$ states, the two-body correlation
function $f_{s}(r)$ is obtained by numerically integrating
 Eq. (\ref{Dif}). Using this two-body correlation function we can
determine the effective pair potential $w_{s}(r)$ as a function of
interatomic distance from Eq. (\ref{W_{r12}}). Finally, the
two-body energy $E_2$ and the total energy of system can be
calculated.

\section{Results and Discussion \label{Res}}
We have calculated some ground state properties of the polarized
liquid $^{3}He$ at zero temperature with the Lennard-Jones
\cite{LJ} and Aziz \cite{Aziz1,Aziz2} pair potentials using the
spin-dependent correlation function. Our results are as follows.

\begin{figure}
\includegraphics[height=2.5in]{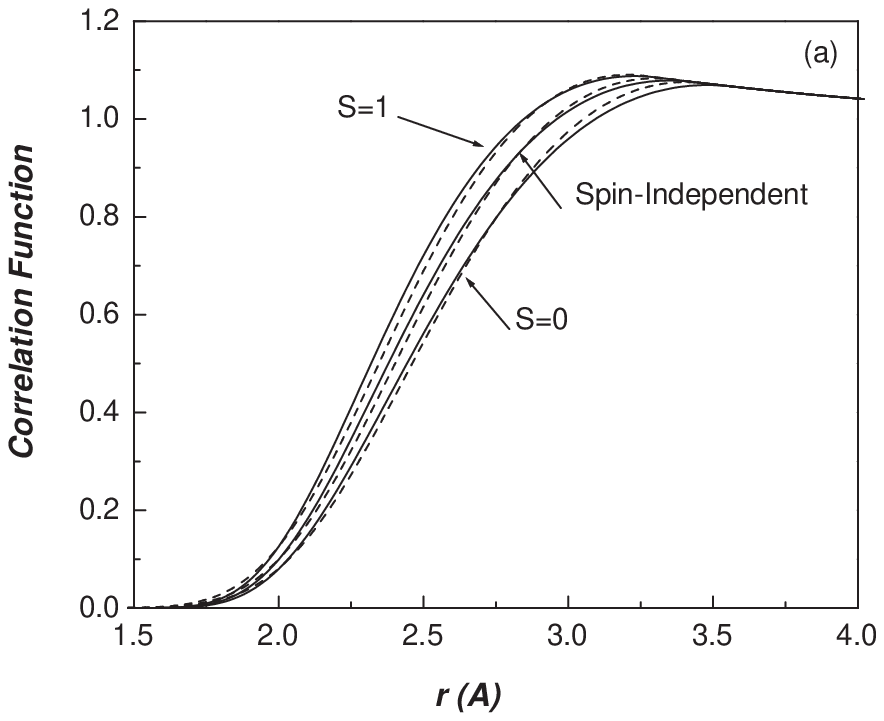}
\includegraphics[height=2.5in]{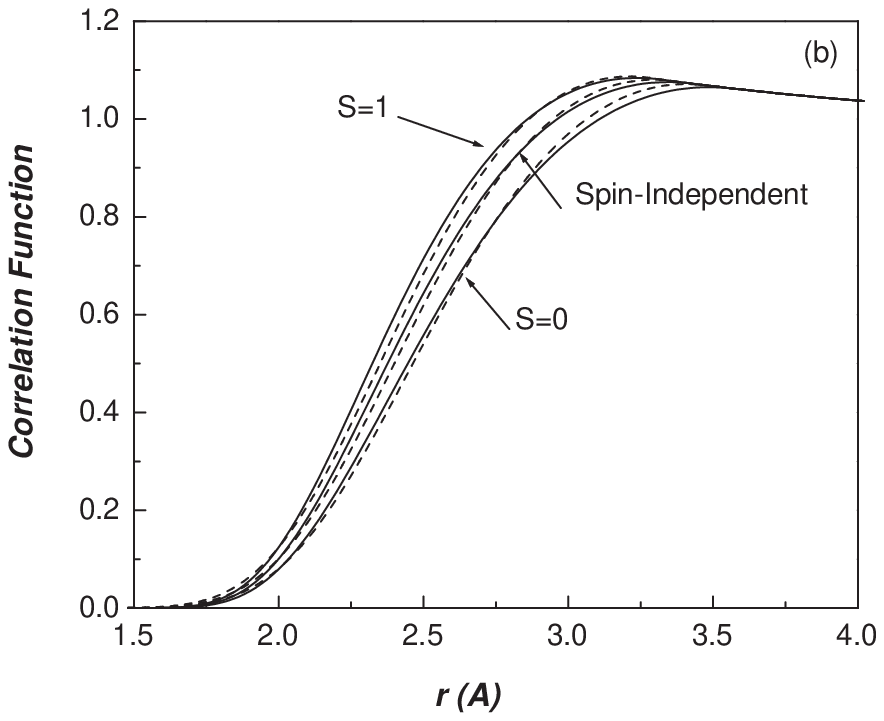}
\includegraphics[height=2.5in]{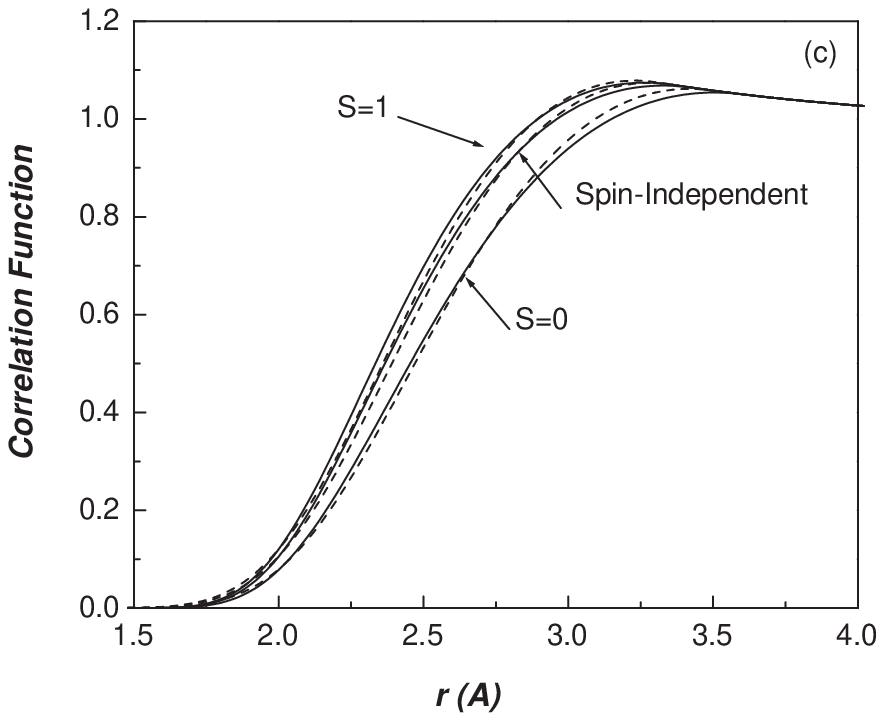}
\includegraphics[height=2.5in]{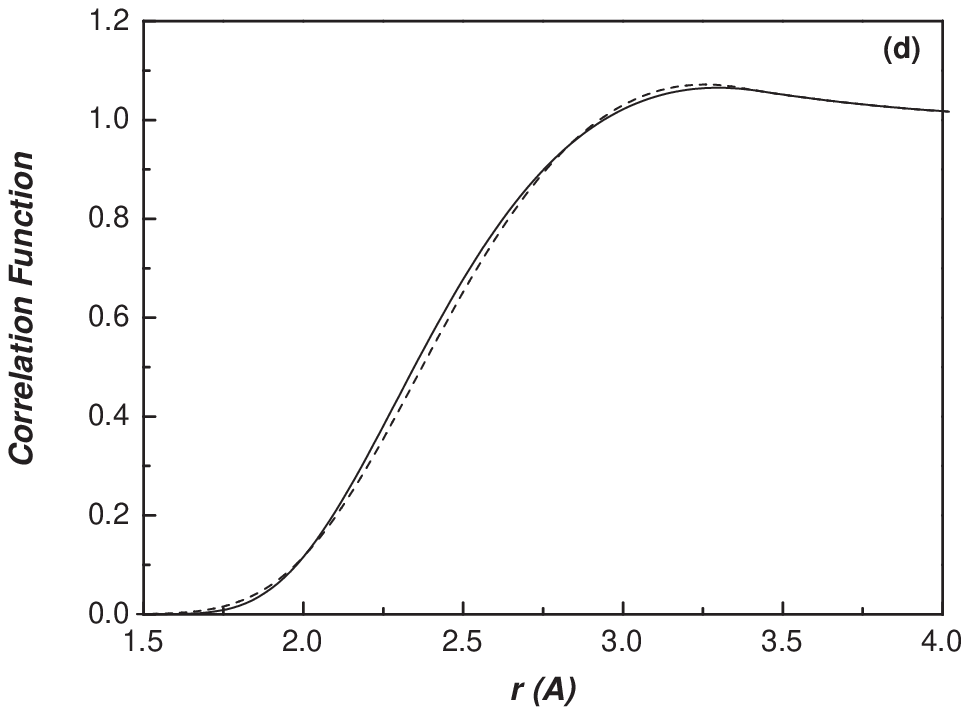}
 \caption{The correlation function with the Aziz (dashed curves)
 and Lennard-Jones (full curves) pair potentials in
the case of spin-dependent at $s=0$ and $s=1$ states for $\xi=0.0$
(a), $\xi=0.33$ (b), $\xi=0.66$ (c) and $\xi=1.0$ (d). Our results
for the spin-independent correlation function are also presented
for comparison.
  } \label{Cor}
\end{figure}

The spin-dependent two-body correlation functions at $s=0$ state
and $s=1$ states for different values of spin asymmetry parameter
($\xi$) are shown in Fig. \ref{Cor}. These figures show that the
correlation function at $s=1$ state heals to pauli function,
$h(r)$, more rapidly than $s=0$ state. Therefore, the $s=1$ state
has a shorter correlation length with respect to $s=0$ state. For
large values of $r$, $f_{0}(r)$ and $f_{1}(r)$ have the same
values and therefore, the spin-dependent part of correlation
operator (Eq. \ref{Corrf}) is vanished. In these figures, the
spin-independent two-body correlation function are also plotted
for comparison. It is seen that the spin-dependent correlation
function differs from spin-independent correlation function,
except for the fully polarized matter ($\xi =1.0$). This is due to
the fact that for fully polarized matter, there is only $s=1$
state. From Fig. \ref{Cor}, we can see that the correlation
functions with the Lennard-Jones and Aziz potentials are nearly
identical.

\begin{figure}
\includegraphics[height=2.5in]{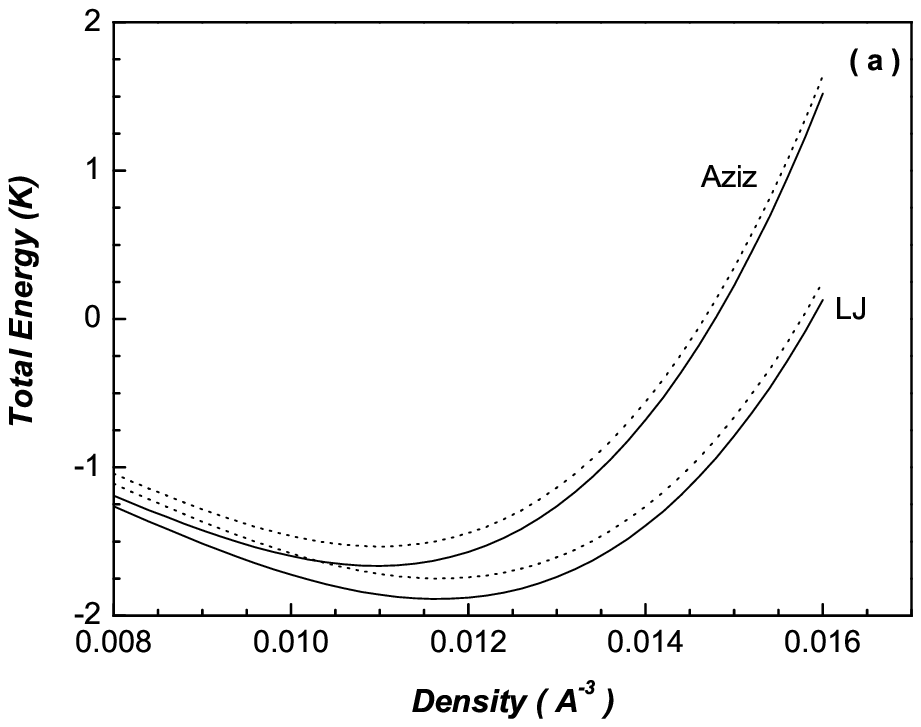}
\includegraphics[height=2.5in]{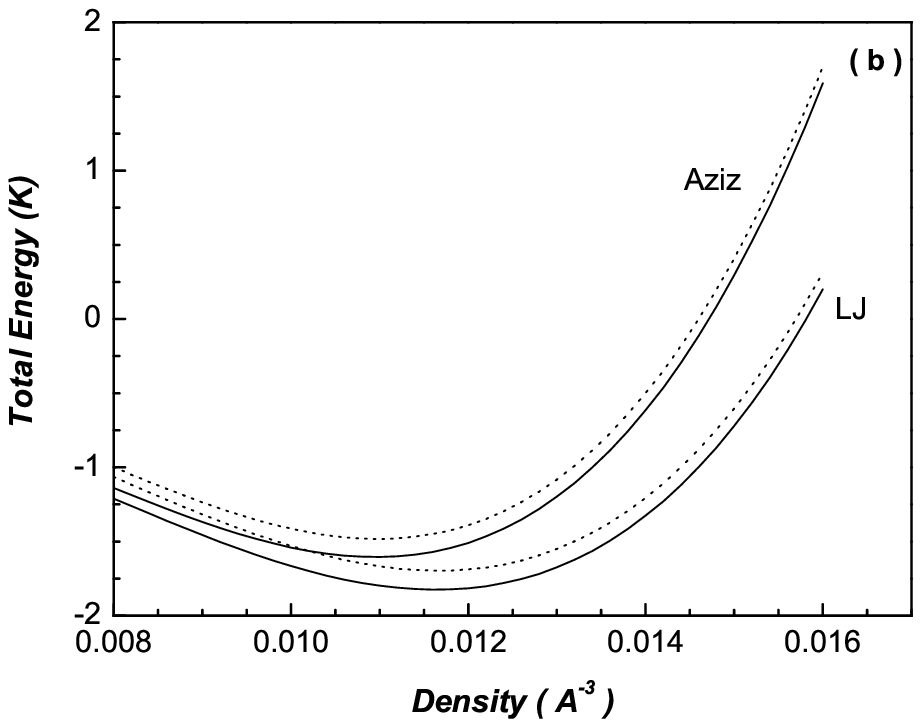}
\includegraphics[height=2.5in]{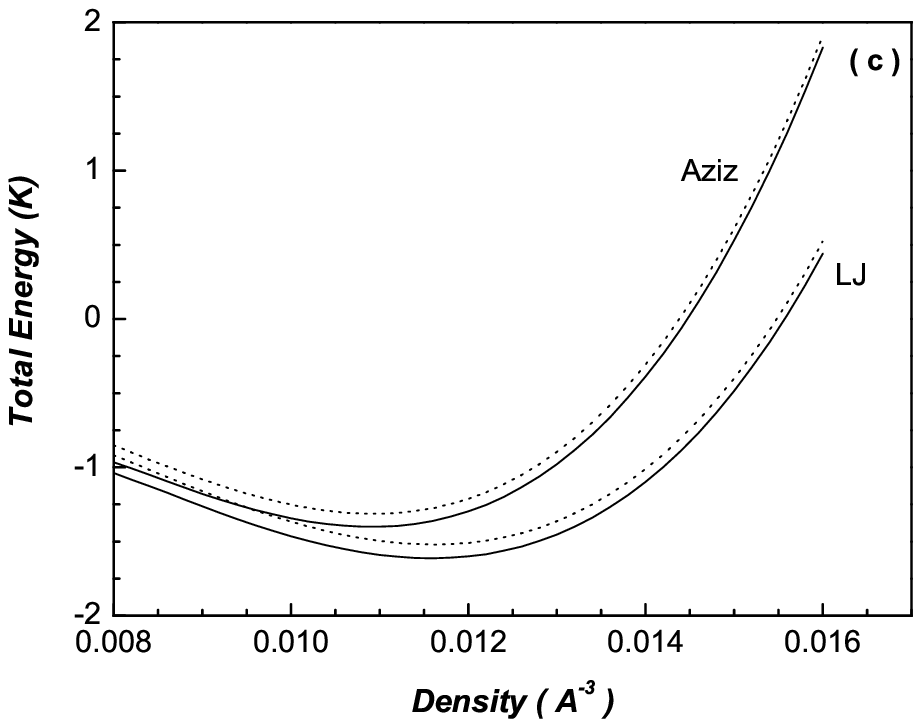}
\includegraphics[height=2.5in]{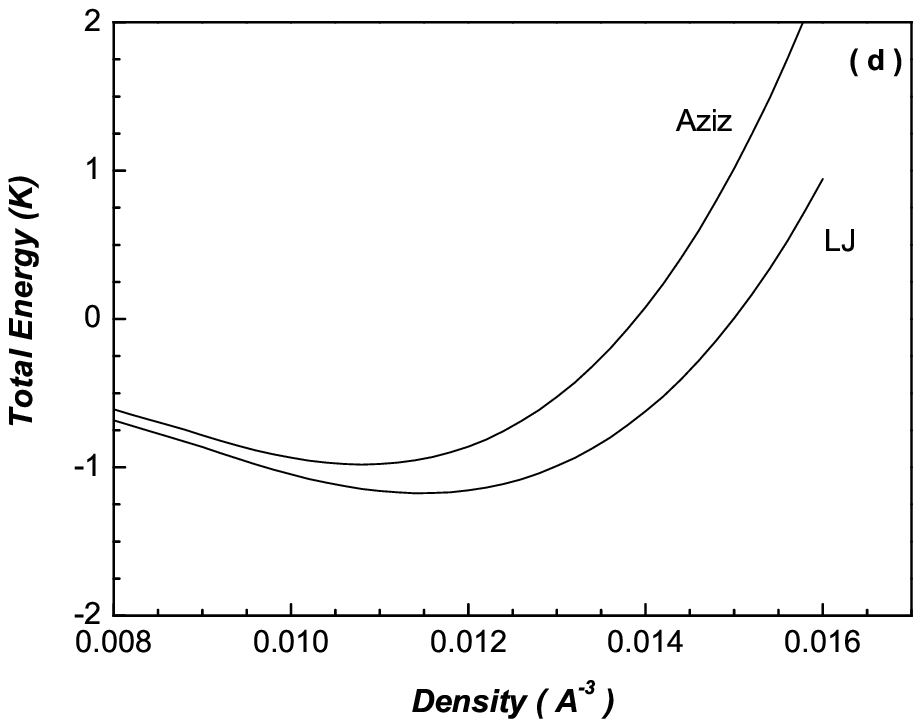}
 \caption{Our results for the total energy of the
 polarized liquid $^3He$ with the Aziz
 and Lennard-Jones (LJ) pair potentials
  in the case of spin-dependent (full curve)
 and spin-independent (dotted curve) correlation functions
  for $\xi=0.0$ (a),
$\xi=0.33$ (b), $\xi=0.66$ (c) and $\xi=1.0$ (d).
  } \label{Tot}
\end{figure}

In Fig. \ref{Tot}, we have shown the total energy of polarized
liquid $^3He$ versus number density calculated both with the
spin-dependent correlation and the spin-independent correlation at
different values of spin asymmetry parameter $\xi$. We can see
that the total energy increases by increasing $\xi$. Fig.
\ref{Tot} indicates that in the spin-dependent case, the total
energy of the liquid $^{3}He$ is lower than the spin-independent
case. It is also seen that for all values of $\xi$, the energy
curve has a minimum which shows the existence of a bound state for
this system.  It is shown that the difference between the energies
of spin-dependent case and spin-independent case decreases by
increasing $\xi$ and it becomes zero as $\xi$ approaches to one.
It is seen that for all values of the density and spin asymmetry
parameter, the total energy with the Aziz pair potential is grater
than that of the Lennard-Jones pair potential.

\begin{figure}
\includegraphics[height=2.5in]{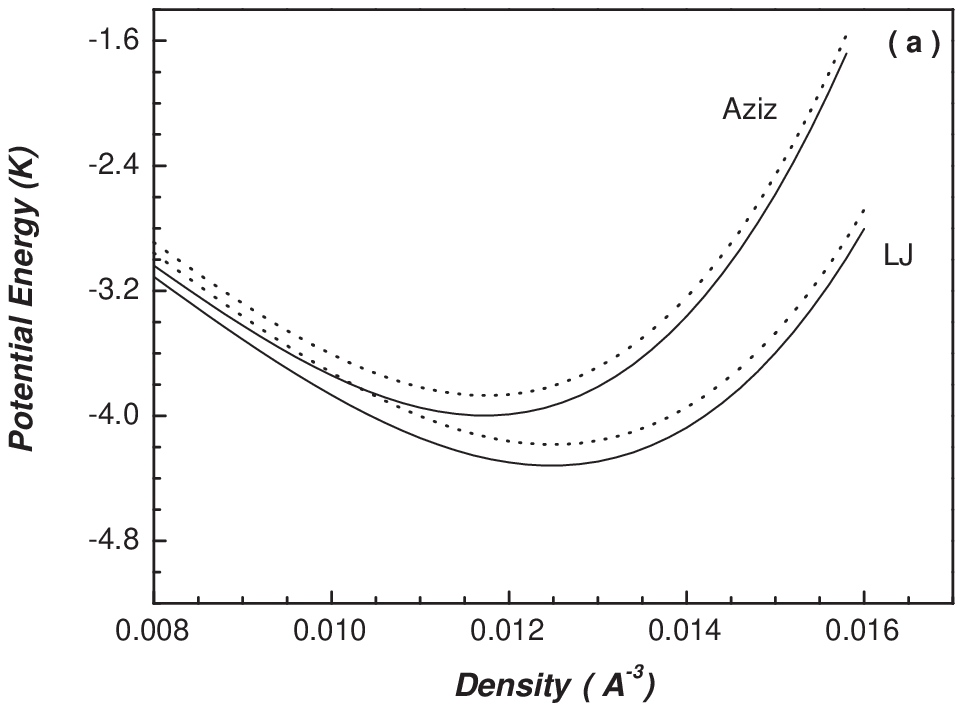}
\includegraphics[height=2.5in]{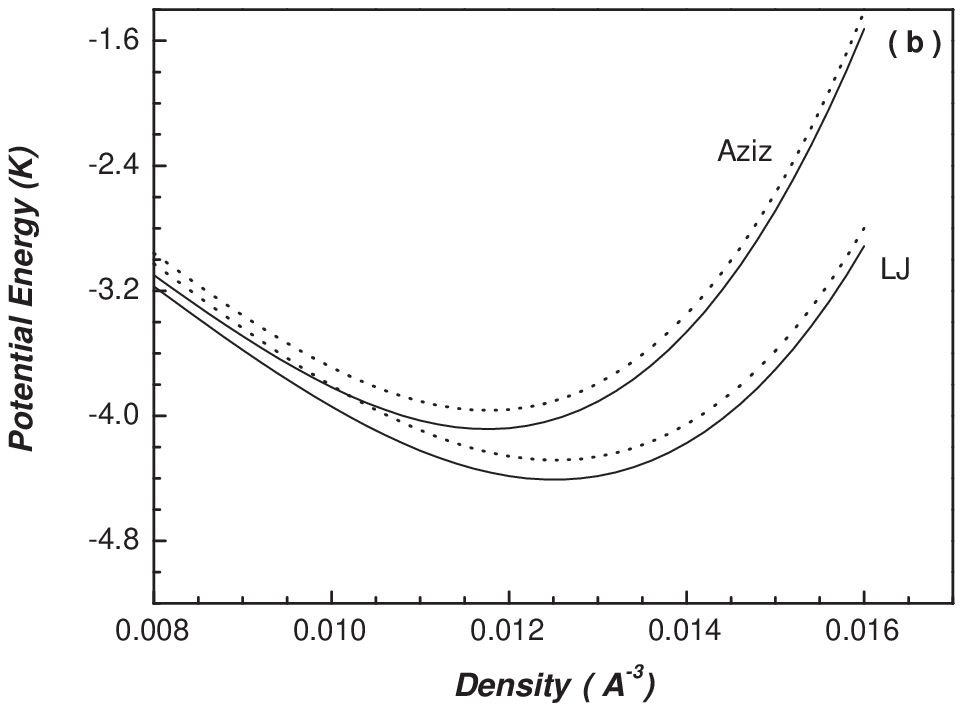}
\includegraphics[height=2.5in]{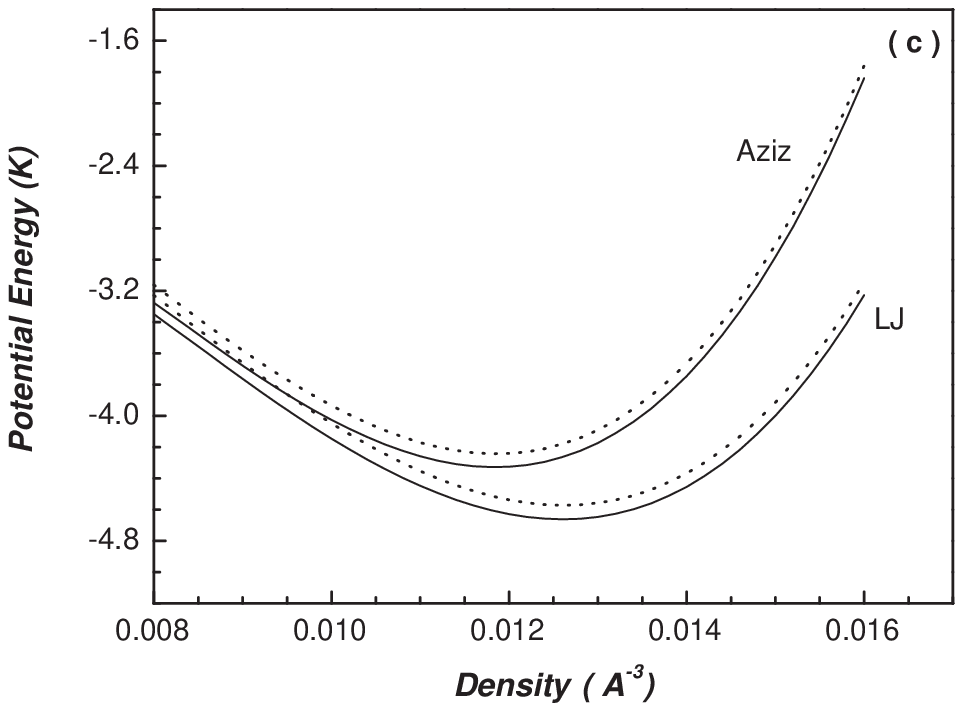}
\includegraphics[height=2.5in]{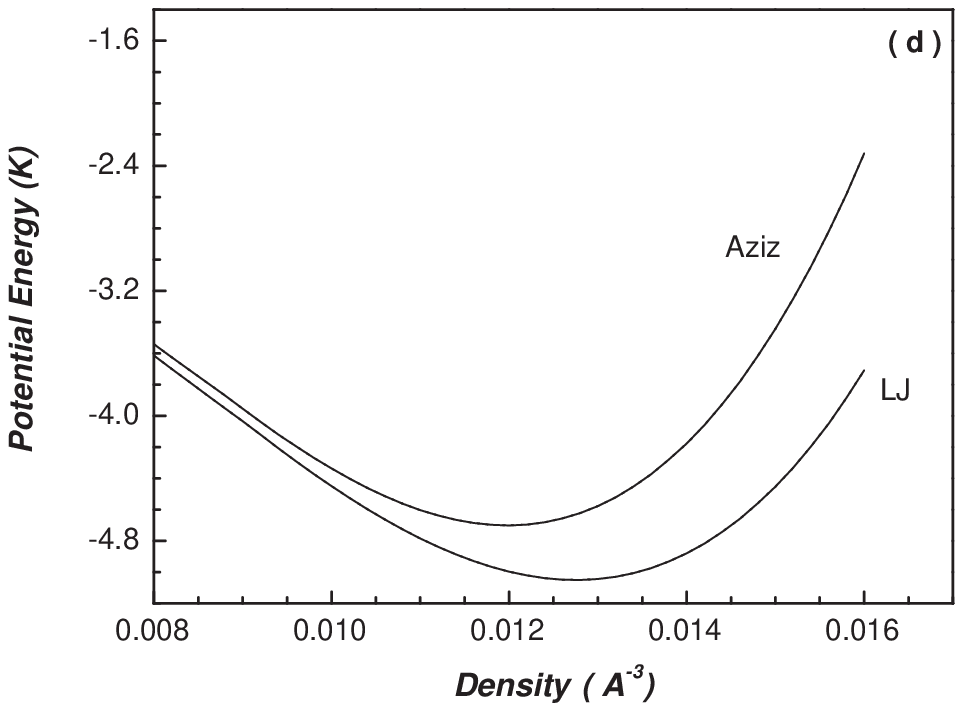}
 \caption{As Fig. 2, but
for the potential energy of the polarized liquid $^3He$.}
\label{Pot}
\end{figure}

The potential energy of the polarized liquid $^{3}He$ for
different values of $\xi$ are presented in Fig. \ref{Pot}, for
spin-dependent and spin-independent cases. This figure indicates
that the potential energy decreases by increasing the
polarization. According to the above results, we can conclude that
the increasing of kinetic energy dominates and this leads to the
increasing of total energy by increasing $\xi$. Fig. \ref{Pot}
shows that the potential energy in the spin-dependent case has
lower values with respect to the spin-independent case. It is also
seen that the difference between the potential energies of
spin-dependent and spin-independent cases decrease by increasing
$\xi$. We see that the potential energies with the Aziz and the
Lennard-Jones pair potentials are different. This difference
increases by increasing the density.

\begin{figure}
\includegraphics[height=2.5in]{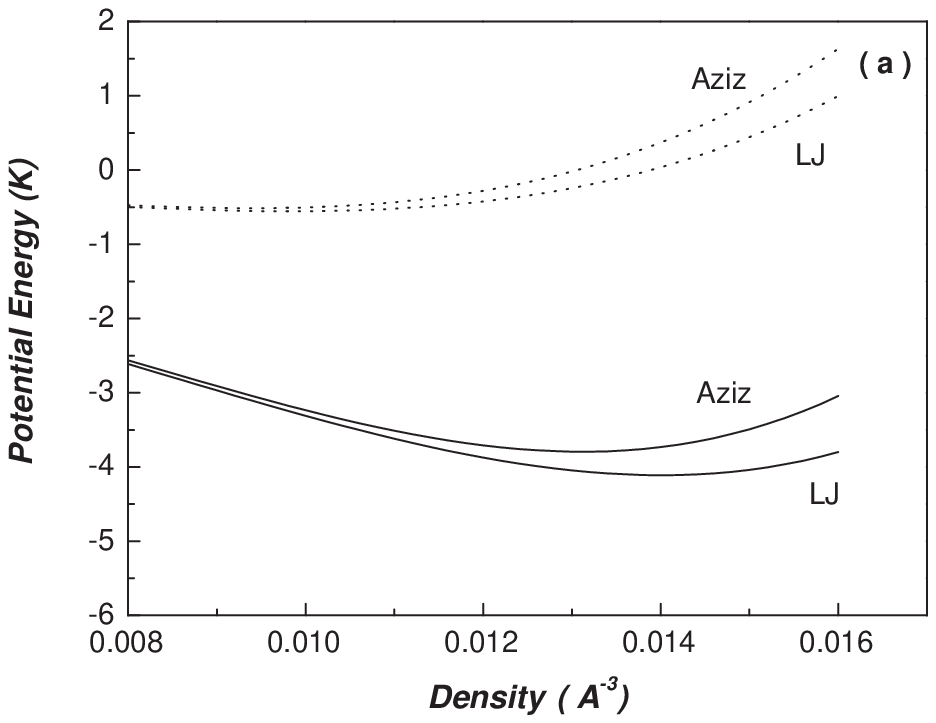}
\includegraphics[height=2.5in]{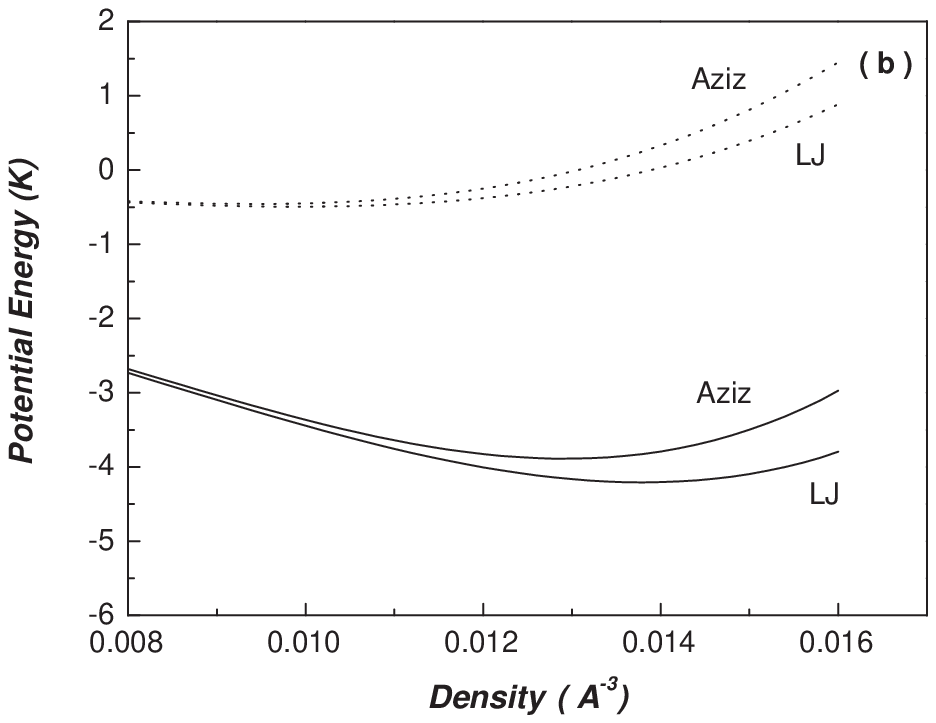}
\includegraphics[height=2.5in]{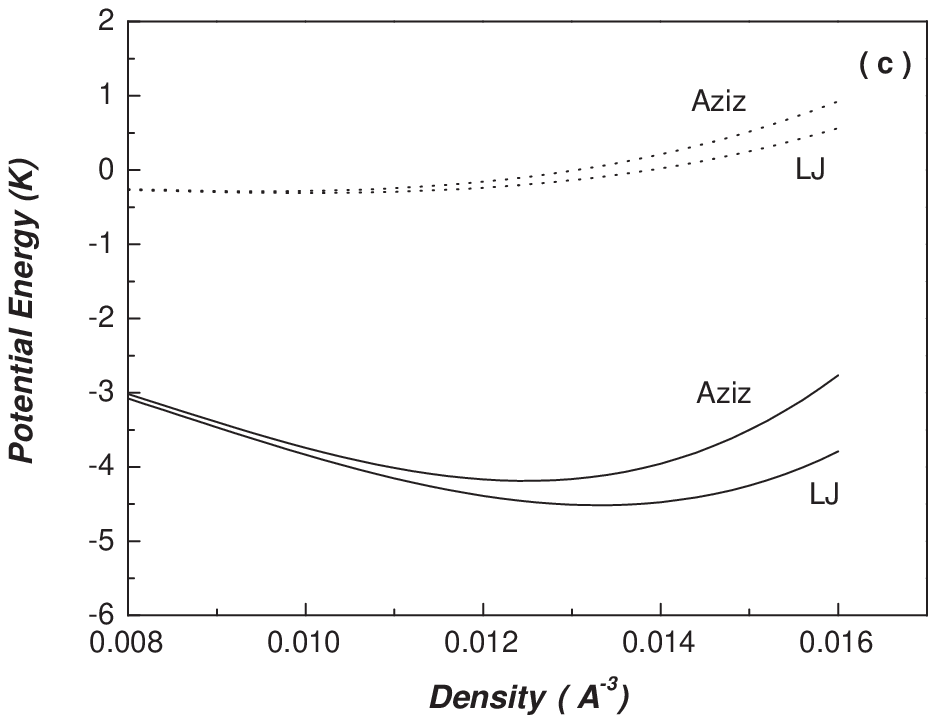}
\includegraphics[height=2.5in]{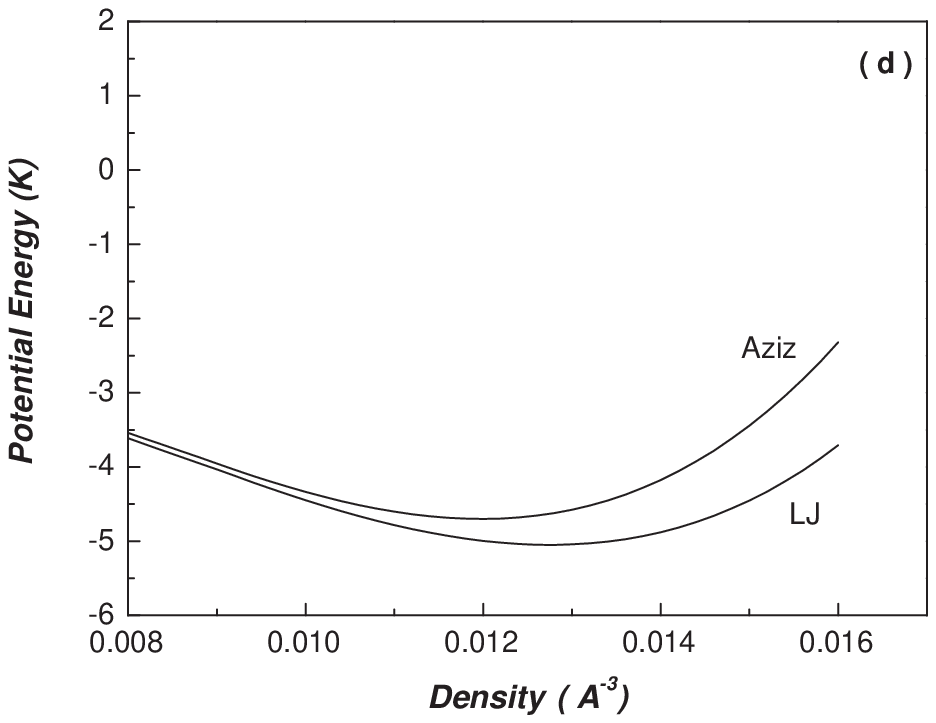}
 \caption{Our results for the potential energy of the polarized liquid $^3He$
 with the Aziz and Lennard-Jones (LJ) pair potentials
  at $s=0$ (dotted curve) and $s=1$ (full curve) states for $\xi=0.0$ (a),
$\xi=0.33$ (b), $\xi=0.66$ (c) and $\xi=1.0$ (d).
  } \label{Potspin}
\end{figure}

 In Fig \ref{Potspin}, the potential energies of
$s=0$ and $s=1$ states for different values of $\xi$ are compared.
We have seen that the potential energy at $s=1$ state is lower
than at $s=0$ state. It can be concluded that the spin-triplet
state has the main contribution in the potential energy of
polarized liquid $^{3}He$. It is also seen that the potential
energy of $s=0$ ($s=1$) state increases (decreases) by increasing
$\xi$.
 For $s=0$ state, we can see that at low densities,
 the potential energies
 with the Lennard-Jones and Aziz pair potentials are nearly identical.
 However, for $s=1$ state and high densities, the
difference between these potential energies becomes appreciable.

\begin{figure}
\includegraphics[height=3.7in]{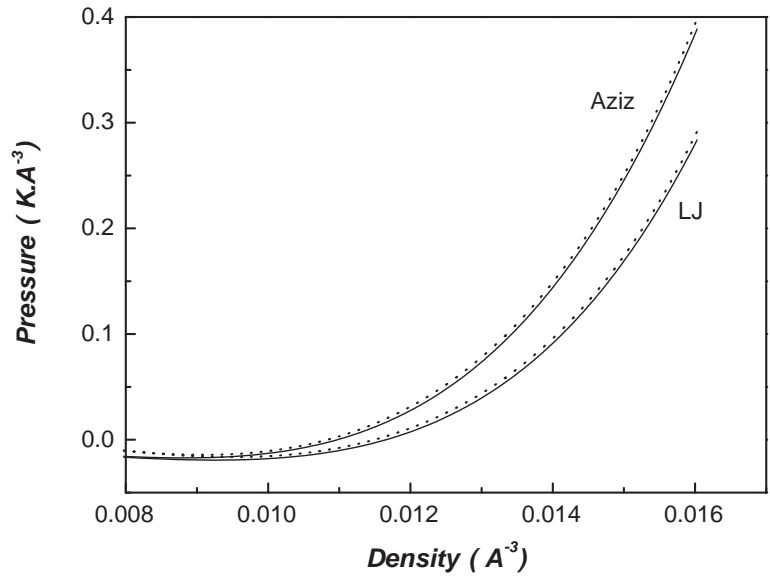}
\caption{The equation of state of the fully polarized (dotted
curve) and unpolarized (full curve) liquid $^3He$ with the Aziz
 and Lennard-Jones (LJ) pair potentials.}
\label{pressure}
\end{figure}

The equation of state of polarized liquid $^{3}He$, $P(\rho,
\xi$), can be obtained using
\begin{eqnarray}
      P(\rho, \xi)= \rho^{2} \frac{\partial E(\rho, \xi)}{\partial \rho}
 \end{eqnarray}
In Fig. \ref{pressure}, we have presented the pressure of liquid
$^{3}He$ as a function of the density ($\rho$) for fully polarized
($\xi=1.0$) and unpolarized ($\xi=0.0$) cases. This figure shows
that for different values of the polarization, the equations of
state of liquid $^{3}He$ are nearly identical.
 From Fig. \ref{pressure}, it is seen that
 for both $\xi=1.0$ and $\xi=0.0$,
 the equation of state with the Aziz pair potential is
stiffer than that of the Lennard-Jones pair potential.

\section{Summary and Conclusion}
We have considered a system consisting of Helium atoms $(^{3}He)$
with asymmetrical spin configuration and derived the two-body term
in the cluster expansion of the energy functional by employing
spin-dependent correlation function. Then, we have minimized the
two-body energy term under the normalization constraint and
obtained the Euler-Lagrange differential equation. By numerically
solving this differential equation, we have computed the
correlation function and then calculated the other properties of
this system with the Lennard-Jones and Aziz pair potentials. It is
shown that for the two different spin-singlet and spin-triplet
states, the correlation functions are different from each other.
Our results show that the introduction of the spin-dependent term
in the correlation operator reduces the total energy of system by
about $10\%$. It is also shown that the total energy increases by
increasing the polarization. The difference between the energies
of the spin-dependent and spin-independent cases decreases by
increasing the polarization. We have seen that, the potential
energy of these states have a remarkable difference. It is shown
that the main contribution of the potential energy comes from
$s=1$ state. Our calculations show that there is a difference
between the results with the Lennard-Jones and Aziz pair
potentials, especially at high densities.

\acknowledgements{ Financial support from the Shiraz University
research council is gratefully acknowledged.}


\newpage

\end{document}